\documentstyle[aps,multicol,prl]{revtex}
\renewcommand{\narrowtext}{\begin{multicols}{2} \global\columnwidth20.5pc}
\renewcommand{\widetext}{\end{multicols} \global\columnwidth42.5pc}

\input{epsf.tex}
\def\inseps#1#2{\def\epsfsize##1##2{#2##1} \centerline{\epsfbox{#1}}}

\def \attn #1 {{\sl $\bullet$ #1 $\bullet$}}
\begin{document}
\draft

\title{Quantum Phases of Vortices in Rotating Bose-Einstein
Condensates}
\author{N. R. Cooper$^{(1)}$, N. K. Wilkin$^{(2)}$, J. M. F. Gunn$^{(2)}$}
\address{(1) Theory of Condensed Matter Group, Cavendish Laboratory, 
Madingley Road, Cambridge, CB3 0HE,
United Kingdom.}
\address{(2) School of Physics and Astronomy, University of Birmingham, Edgbaston,
Birmingham, B15 2TT, United Kingdom.}
\date{June 29, 2001}

\maketitle

\begin{abstract}

We investigate the groundstates of weakly interacting bosons in a
rotating trap as a function of the number of bosons, $N$, and the
average number of vortices, $N_V$. We identify the filling fraction
$\nu\equiv N/N_V$ as the parameter controlling the nature of these
states. We present results indicating that, as a function of $\nu$,
there is a zero temperature {\it phase transition} between a
triangular vortex lattice phase, and strongly-correlated vortex liquid
phases. The vortex liquid phases appear to be the
Read-Rezayi parafermion states.

\end{abstract}


\narrowtext

A fundamental characteristic of condensed Bose systems is their
response to rotation\cite{Leggett}. A transition to a ``normal'' phase
might be expected at sufficiently high angular velocities, $\omega$,
of the container (or trap) by loose analogy with a superconductor in a
magnetic field. At zero temperature this phase would constitute a
novel {\it uncondensed} ground state. Such a regime is entered when
the vortex cores start to overlap. The corresponding value of $\omega$
is unattainable with bulk $^4$He, but may be achievable in the very
dilute degenerate atomic gases initially explored in
Ref.~\cite{WilkinGS98}, and studied extensively in 
Refs.~\cite{GP,various,WilkinG00,CooperW99,ViefersHR00,JacksonKMR01,Liu00},
Apart from the identification\cite{WilkinG00} of the Laughlin state as
the ground state at sufficiently high $\omega$, work on the most
interesting regime of large numbers of vortices has been restricted to
either mean field theory\cite{GP} or exact
diagonalisation\cite{WilkinG00,CooperW99,ViefersHR00,JacksonKMR01}.
These two approaches have exhibited apparently contradictory pictures.
Within Gross-Pitaevskii (GP) mean-field theory, the groundstates are
vortex lattices (distorted by the confinement), with broken rotational
symmetry\cite{GP}.  On the other hand, exact diagonalisations have
identified groundstates which do {\it not} have crystalline
correlations of vortex locations\cite{WilkinG00}; they are strongly
correlated vortex-liquids, closely related to incompressible liquid
states responsible for the fractional quantum Hall
effect\cite{WilkinG00,CooperW99,ViefersHR00}.

Here we present results of extensive exact diagonalisations (EDs) that
elucidate the relationship between these two pictures. By using a
periodic geometry, we have been able to study systems containing many
vortices up to boson densities far in excess of previous EDs.  Our
results indicate that both crystalline and liquid phases of vortices
exist. A clean distinction between these phases can only be made for a
large number of vortices.  In this limit, we argue that there is a
zero-temperature phase transition as a function of the ``filling
fraction'', $\nu\equiv N/N_V$, the ratio of the number of bosons, $N$,
to the average number of vortices, $N_V$. For large $\nu$, the
groundstate is a vortex lattice (characterised by broken
translational/rotational symmetry). For small $\nu$ the groundstates
are strongly-correlated vortex liquids.  We find that the
vortex-liquid groundstates are related to the Read-Rezayi
``parafermion'' states\cite{ReadR99} that were introduced in the
context of fractional quantum Hall systems.

In a frame of reference rotating with angular velocity
$\omega\hat{\bbox{z}}$, the Hamiltonian for a particle of mass $m$ in an
(isotropic) harmonic trap of natural frequency $\omega_0$ is
\begin{eqnarray*}
H_\omega &  = & \frac{\bbox{p}^2}{2m} + \frac{1}{2} m\omega_0^2\bbox{r}^2- \omega \hat{\bbox{z}}\cdot\bbox{r}\times\bbox{p}\\
 &  = & \frac{(\bbox{p}-m\omega\bbox{\hat{z}}\times\bbox{r})^2}{2m} + \frac{1}{2} m\left[(\omega_0^2-\omega^2)(x^2+y^2) + \omega_0^2 z^2\right] .
\label{eq:ham}
\end{eqnarray*}
The second form indicates the equivalence to the Hamiltonian of a
particle of charge $q^*$ experiencing an effective magnetic field
$\bbox{B}^* = \nabla\times (m\omega \hat{\bbox{z}}\times\bbox{r}/q^*)
= (2m \omega/q^*)\hat{\bbox{z}}$ (the particle also feels a reduced
$xy$-confinement). Of particular importance to our discussion is the
average filling fraction, $\nu$, for the bosons in this effective
magnetic field.  For $N$ bosons spread over an area $A$, one finds
\begin{equation}
\nu  \equiv   \frac{N}{A} \frac{h}{q^*B^*} = \frac{N}{A} \frac{h}{2m\omega}\\
  =   \frac{N}{N_V},
\label{eq:nu}
\end{equation}
where $N_V$ is the {\it average number of vortices}. For large $N_V$
the vortex density is approximately uniform, and $N_V = (2 m \omega
A)/h$, or (equivalently) $N_V = 2L/N$\cite{footnote_nv}, where $L$ is
the total angular momentum in units of $\hbar$.

We now introduce repulsive interactions\cite{DalfovoGP99}
\begin{equation}
V = g \sum_{i< j=1}^N \delta(\bbox{r}_i-\bbox{r}_j),
\label{eq:v}
\end{equation}
with $g = 4\pi\hbar^2a/m$, chosen to give the correct $s$-wave
scattering length $a$.  Throughout this work, we make use of the limit
of weak interactions formulated in Ref.\onlinecite{WilkinGS98}. For $g
\ll \hbar\omega_0 \bar a^3$, with $\bar a$ the interparticle spacing,
the bosons are restricted to single particle states in the lowest
Landau level, and lowest oscillator state of $\bbox{z}$.  For
$\omega\sim \omega_0$, the repulsive interactions give rise to the
appearance of rotating (vortex)
groundstates\cite{GP,various,WilkinG00,CooperW99,ViefersHR00,JacksonKMR01,Liu00}.

GP theory\cite{GP} takes account of interactions by finding the
fully-condensed state that minimises the total energy.  In
EDs\cite{WilkinG00,CooperW99,ViefersHR00}, the groundstate is found by
diagonalising (\ref{eq:v}) within the set of all states of $N$ bosons
with fixed total angular momentum $L$ ($L$ is conserved by
interactions). An important distinction between these two approaches
is that the GP groundstates exhibit broken rotational
symmetry\cite{GP}, while the ED groundstate is necessarily an
eigenstate of angular momentum, $L$.  However, by performing EDs on
large numbers of bosons (up to $N=30$ at $L\sim 2N$), we find that, as
$N$ becomes large at fixed $L/N$, a macroscopic number of
quasi-degeneracies appear between states with different $L$. This
signals the emergence of broken rotational symmetry.  Indeed, it
appears from these and other\cite{JacksonKMR01,Liu00} studies that as
$N$ becomes large for fixed $L/N$, there is a crossover to a regime in
which GP theory is essentially correct.  We believe that this
crossover is related to the phase transition, discussed in detail
below, between vortex liquids at small $\nu$, and a vortex lattice at
large $\nu$. Applying a simple Lindemann criterion\cite{RozhkovS96},
one finds that a triangular vortex lattice is unstable to quantum
fluctuations for $\nu \lesssim 14$.  The crossover to GP behaviour for
increasing $N$ at fixed $L/N$ is the remnant of this phase transition
in a system with a finite number of vortices $N_V= 2L/N$.

To investigate in detail the dependence of the groundstate on $\nu$,
we have conducted extensive (Lanczos) diagonalisations in a {\it
toroidal} geometry\cite{YoshiokaHL83}.  This periodic geometry
represents the bulk of a system containing a large number of vortices.
We consider a torus of sides $a$ and $b$.  There are then $N_V=
(2m\omega ab)/h=ab(q^*B^*/h)$ vortices, which is the number of
single-particle states on the torus in the lowest Landau
level\cite{YoshiokaHL83}, and hence an integer. Thus, both $N$ and
$N_V$ are integers, and $\nu\equiv N/N_V$ is a rational fraction.
Finally, we classify all states by the
Haldane momentum\cite{haldanemtm}, which runs over a Brillouin zone
containing $\bar{N}^2$ points, where $\bar N$ is the greatest common
divisor of $N$ and $N_V$. In the following we shall refer to the $x$
and $y$ momenta by the dimensionless vector, $(K_x,K_y)$, using units
of $(2\pi\hbar/a)$ and $(2\pi\hbar/b)$. We report only positive values
of $K_x,K_y$ up to the Brillouin zone boundary [states at $(\pm
K_x,\pm K_y)$ are degenerate by symmetry].  
We also choose to measure energies in units of $g/(\sqrt{4\pi}\ell^3)$,
where  $\ell \equiv \sqrt{\hbar/(q^*B^*)}= \sqrt{\hbar/(2m\omega_0)}$
is the magnetic length at $\omega=\omega_0$.

We start by applying Gross-Pitaevskii theory\cite{GP} on the
torus.  In general, the GP groundstate is a {\it vortex lattice}, with
broken translational symmetry: the wavefunction is not an eigenstate
of the Haldane momentum, but has weight at a set of reciprocal lattice
vectors (RLVs).  While the symmetry of the lattice depends, in
general, on $N_V$ and the  aspect ratio $a/b$,
the absolute minimum of energy
is always obtained for a triangular vortex lattice\cite{KleinerRA64}.

\begin{figure}
\inseps{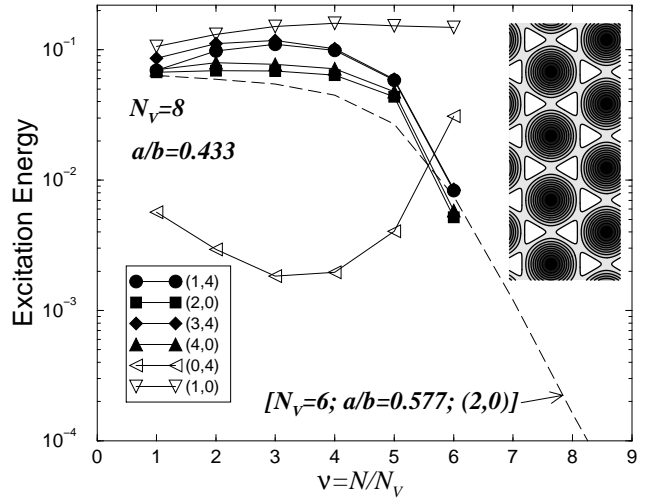}{0.5} 
\caption{Solid lines: Excitation energies at momenta measured relative
to the groundstate, for $N_V=8$,  $a/b=\sqrt 3 /4$ (inset
shows the GP groundstate: dark = low boson density).  The excitation
energies at the RLVs of the triangular lattice
(filled symbols) collapse at $\nu\sim 6$, signalling the onset of a
groundstate quasi-degeneracy; all other momenta retain non-zero
excitation energies (two such momenta are shown as open
symbols). Dashed Line: The excitation energy at one RLV $(2,0)$ for
$N_V=6$ and $a/b=1/\sqrt{3}$, showing that the collapse
at $\nu\sim 6$ initiates an exponential decrease with $\nu$.}
\label{fig:NV8ab0.433} 
\end{figure}

In ED studies, the groundstate is necessarily an eigenstate of the
Haldane momentum. The signature of translational symmetry breaking is
the development of quasi-degenerate levels at the set of momenta given
by the RLVs of the broken symmetry lattice\cite{RezayiHY99}. To search
for such degeneracies, we show in Fig.~\ref{fig:NV8ab0.433} the
evolution with $\nu$ of the excitation energies for $N_V=8$ vortices
at an aspect ratio ($a/b=\sqrt 3 /4$) for which the GP groundstate is
a triangular lattice.
A collapse of the excitation energies at the RLVs of a triangular
lattice is observed at $\nu\sim 6$. Similar plots for $N_V=4,6$
indicate that the excitation energies at RLVs fall exponentially with
$\nu$ for $\nu\gtrsim 6$ (shown in Fig.~\ref{fig:NV8ab0.433} for one
RLV for $N_V=6$). For $\nu=15$ at $N_V=6$ the excitation energies are
6 orders of magnitude smaller at the RLVs than at any other momentum.
This strong quasi-degeneracy at the {\it reciprocal lattice vectors of
the lattice formed in GP theory} indicates a strong tendency to broken
translational symmetry\cite{RezayiHY99}.  GP theory accurately
describes the states at large values of $\nu$.

We view the collapse of the excitation gaps at $\nu\sim 6$ as an
indication, in this finite-size system, of a true phase-transition
from translationally-invariant ``vortex-liquid'' phases, to a
(triangular) vortex lattice.  The phase transition is rounded due to
the finite number of vortices, and becomes sharper for larger $N_V$
(over the range of $N_V$ we can study). Note that we have chosen
aspect ratios that are commensurate with a triangular lattice, which
is likely to help stabilise the vortex lattice. Similar plots at other
aspect ratios show transitions to a vortex lattice at larger values of
$\nu$ (up to $\nu_c\sim 15$ for $N_V=4$).  One should therefore view
$\nu_c\sim 6$ as a {\it lower bound} on the critical value of $\nu$ at
which the transition occurs. Our numerical results are consistent with
a transition in the vicinity of $\nu_c\sim 10\pm 5$.

\begin{figure}
\inseps{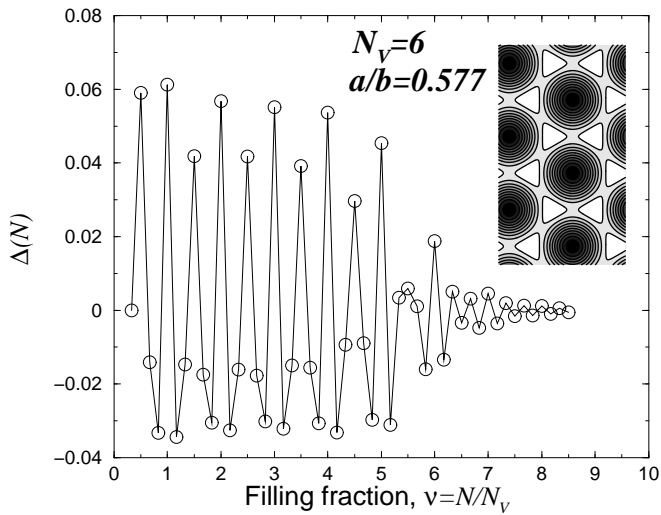}{0.5}
\caption{Energy gap (\ref{eq:gap}) as a function of $\nu$ for $N_V=6$
vortices, at $a/b=1/\sqrt 3$.  Upward spikes signal values of $\nu$
for which the groundstate is incompressible. The collapse of the gaps
at $\nu\sim 6$ indicates the transition to the vortex lattice
phase. (Inset shows the density of the GP groundstate.)}
\label{fig:NV6delsq} \end{figure}

We now turn to discuss the vortex liquids at $\nu\lesssim 6$.  In this
regime, we find incompressible liquid states similar to those in
fractional quantum Hall systems.  Some of these incompressible states
can be accounted for by the use of a composite fermion construction
that has previously been shown to describe accurately ED results on
small systems in the disk geometry\cite{CooperW99}.  In the present
uniform geometry, this theory predicts a sequence of incompressible
states at $\nu= \frac{\nu_{cf}}{\nu_{cf}+1} = \frac{1}{2},
\frac{2}{3}, \frac{3}{4}, \frac{4}{5}\ldots \frac{5}{4}, \frac{4}{3},
\frac{3}{2}, 2,\infty$, which is a bosonic version of the Jain
sequence of fractional quantum Hall states\cite{jainoriginal}.  Many
of the strongest incompressible states we find cannot be accounted for
in this way. In particular, the largest (finite) value in the
composite fermion sequence is $\nu=2$, while the transition to a
vortex lattice does not occur until $\nu=6$.  To
investigate the liquid states in this regime, we plot in
Fig.~\ref{fig:NV6delsq} the energy gaps as a function of $\nu$ for
$N_V=6$ vortices.  The energy gap is related to the discontinuity in
the chemical potential. To minimise finite size effects, we define the
gap $\Delta$ by
\begin{equation} \Delta(N) \equiv N \left[ \frac{E(N+1)}{N+1} +
\frac{E(N-1)}{N-1} - 2 \frac{E(N)}{N} \right] ,
 \label{eq:gap}
\end{equation} 
which reduces to the standard definition as $N\rightarrow \infty$.

As well as the Laughlin state at $\nu=\frac{1}{2}$\cite{WilkinG00},
incompressible states appear clearly in Fig.~\ref{fig:NV6delsq} at
$\nu=1,\frac{3}{2},2,\frac{5}{2},3,\frac{7}{2},4,\frac{9}{2},5,6$ (the
loss of gaps  for $\nu\gtrsim 6$ is another indication of the
transition to the vortex lattice, perhaps re-entrant around $\nu=6$).
It is not immediately apparent how to construct incompressible states
for this sequence of $\nu$.
One possibility is that the vortices themselves are forming Laughlin
states. This would provide a set of states with vortex filling
fraction $\nu_V=\frac{1}{2},\frac{1}{4},\frac{1}{6}\ldots$\cite{Stern94}, and hence
$\nu=1/\nu_V=2,4,6\ldots$.  However, this construction does not
account for states at
$\nu=1,\frac{3}{2},\frac{5}{2},3,\frac{7}{2},\frac{9}{2},5$\cite{footnote_pairedvortices}. 
Moreover,
the trial wavefunctions of this form that we have tested (on a disc)
have high interaction energies: while they keep the vortices apart,
they do not introduce favourable correlations between the
bosons. States of this type are likely to describe systems in which
the underlying interactions can be described as repulsive two-body
forces between vortices\cite{Stern94,Horovitz95,RozhkovS96}.  They do
not provide an accurate description in the present situation, where the
interactions cannot be represented by pairwise vortex
interactions\cite{TesanovicX91}.

A clue to the nature of the incompressible vortex liquid states lies
in the existence of an incompressible state at $\nu=1$, which also
cannot be accounted for in terms of non-interacting composite
fermions.  Rather, this state is well-described\cite{haldaneunpub} by
the Moore-Read (``Pfaffian'') wavefunction\cite{MooreR91}. We find
that the exact groundstate has large overlap with the Moore-Read
state at the Haldane momenta for which it can be
constructed on a torus\cite{pfaffiantorus}.

Motivated by this success, we have compared the incompressible states
at higher integer and half-integer $\nu$ with ``parafermion''
wavefunctions introduced by Read and Rezayi\cite{ReadR99} as
generalisations of the Moore-Read state.  These states may be
represented\cite{CappelliGT01} as a (symmetrised) product of $k$
Laughlin states via
\begin{equation}
\Psi^{(k)}(\{z_i\}) = {\cal S}\left[\prod_{i< j\in A}^{N/k}\!\!(z_i-z_j)^2\prod_{l< m\in B}^{N/k}\!\!(z_l-z_m)^2\ldots\right]
\label{eq:cluster}
\end{equation}
where $z=x+iy$, and we omit the exponential factor of lowest Landau
level states as usual.  The symbol ${\cal S}$ indicates symmetrisation
over all partitions of $N$ particles into sets $A,B,\ldots$ of $N/k$
particles (we assume that $N$ is divisible by $k$).  The cases $k=1$
and $k=2$ correspond to the Laughlin and Moore-Read wavefunctions.
For general $k$, the wavefunction (\ref{eq:cluster}) describes a
system with filling fraction $\nu^{(k)}=N^2/2L= k/2$, and is a zero
energy eigenstate of a $(k+1)$-body version of the repulsion
(\ref{eq:v}).

The Read-Rezayi states provide a consistent interpretation of the
incompressible states in Fig.~\ref{fig:NV6delsq}: they identify the
sequence of incompressible states observed in the EDs
($\nu=\frac{k}{2}$ with integer $k$); they have large overlaps with
the exact wavefunctions, at least up to $\nu=3$ (the largest $\nu$ for
which we have made the comparison). We construct the Read-Rezayi
states on the torus by diagonalising the $(k+1)$-body force-law
directly to find the zero energy eigenstates.  In general, we find
more than one zero energy eigenstate, and recover a total groundstate
degeneracy on a torus of $k+1$, consistent with Ref.\cite{ReadR99}.  The overlap of the exact groundstates of the
two-body force (\ref{eq:v}) with the Read-Rezayi states are given in
Table~\ref{table} for $\nu=\frac{k}{2}$. For comparison, the overlaps
with the GP groundstate are also shown.

In conclusion, we have shown that the groundstates of weakly
interacting bosons in a rotating trap exhibit both vortex lattices and
incompressible vortex liquids.  A clear distinction between these
phases appears for a large number of bosons, $N$, and vortices, $N_V$,
and is controlled by the filling fraction $\nu\equiv N/N_V$. Vortex
liquid phases appear for $\nu\lesssim \nu_c$ and vortex lattices
appear for $\nu\gtrsim\nu_c$. A Lindemann criterion suggests
$\nu_c\sim 14$, while exact diagonalisations 
indicate $\nu_c\gtrsim 6$.  Current experiments\cite{MadisonCWD00}
with $N\sim 10^5$ and $N_V\sim 10$ are deep in the regime in which
the groundstate is a vortex lattice.  Experiments that access the
quantum-melted vortex liquid phases will require specific attention to
small system sizes and high angular momentum. Our results indicate
that novel correlated states emerge in this regime, which are
well-described by the Read-Rezayi parafermion states whose excitations
obey non-abelian statistics\cite{ReadR99}.

The authors are grateful to J. Chalker, M. Dodgson, M. Gaudin,
V. Gurarie, J. Jain, V. Pasquier, N. Read and E. Rezayi for helpful
comments and discussions. This work was partly supported by the Royal
Society and the Nuffield Foundation.

\vskip-0.6cm
\begin{table}
\begin{tabular}{|c|c|c|c|c|}
$k$ & $\nu$ & $(K_x,K_y)\times$degeneracy & $|\langle \Psi^{(k)}
|\Psi\rangle|$ &$|\langle\Psi^{GP}  |\Psi\rangle|$ \\
\hline
\hline
1 & 1/2 \mbox{(Laughlin)}& (0,0)$\times2$    &        1.000    &    0.555    \\
\hline
\hline
2 & 1 \mbox{(Moore-Read)} & (3,3)$\times1$ &        0.982         &    N/W \\
\hline
2 & 1  \mbox{(Moore-Read)} & (3,0)$\times1$  &         0.982      &      0.408 \\
\hline
2 & 1  \mbox{(Moore-Read)} & (0,3)$\times1$ &        0.981       &     0.493  \\
\hline
\hline
3 & 3/2 & (0,0)$\times4$    &        0.967    &        0.234 \\
\hline
\hline
4 & 2 & (0,0)$\times2$ &         0.956         &       0.242\\
\hline
4 & 2 & (3,0)$\times1$ &          0.966      &          N/W \\
\hline
4 & 2 & (0,3)$\times1$ &         0.935     &           N/W\\
\hline
4 & 2 & (3,3)$\times1$    &        0.844    &            0.547\\
\hline
\hline
5 & 5/2 & (0,0)$\times6$    &        0.955    &        0.163
\\
\hline
\hline
6 & 3 & (3,3)$\times2$ &         0.960         &       N/W\\
\hline
6 & 3 &(3,0)$\times2$ &         0.944     &          0.198\\
\hline
6 & 3 &(0,3)$\times2$ &         0.744    &           0.534\\
\hline
6 & 3 &(0,0)$\times1$    &        0.852    &            N/W
\end{tabular}
\caption{Wavefunction comparisons of the exact groundstates of the
2-body force-law at $\nu=k/2$, for $N_V=6$ and
$a/b=1/\sqrt{3}$.  In each case we report: the Haldane momenta at
which Read-Rezayi states exist, with degeneracies; the overlap of
the exact groundstate with the Read-Rezayi state (where there is more than one such state, we report the total overlap within this
set); the overlap of the exact groundstate with the GP groundstate (we
first project the GP state onto each component of momentum; N/W
indicates that the GP groundstate has no weight at this momentum).  }
\label{table}
\end{table}

\widetext

\begin{thebibliography}{10}
\bibitem{footnote_nv}
{F}or small $N_V$ there is a departure from this relation, such that a single
  vortex occurs when $L/N=1$.
\end{thebibliography}
\end{document}